\documentclass{epl}
\title{Mobility gap in intermediate valent TmSe}
\author{M. Dumm\inst{1} \and B. Gorshunov\inst{1,2} \and M.
Dressel\inst{1} \and  T. Matsumura\inst{3}} \institute{
  \inst{1} 1.~Physikalisches Institut, Universit{\"a}t
Stuttgart, Pfaffenwaldring 57, 70550 Stuttgart, Germany\\
  \inst{2} General Physics Institute, Moscow,
Russia\\
  \inst{3} Department of Physics, Graduate School of Science, Tohoku University,
Sendai 980-8578, Japan } 

\begin{document}

\maketitle

\begin{abstract}
The infrared optical conductivity of intermediate valence compound
TmSe reveals clear signatures for hybridization of light $d$- and
heavy $f$-electronic states with $m^*\approx 1.6 m_0$ and
$m^*\approx 16 m_0$, respectively. At moderate and high
temperatures, the metal-like character of the heavy carriers
dominate the low-frequency response while at low temperatures
($T_{\rm N}<T< 100 K$) a gap-like feature is observed in the
conductivity spectra below 10 meV which is assigned to be a
mobility gap due to localization of electrons on local Kondo
singlets, rather than a hybridization gap in the density of
states.
\end{abstract}

\section{Introduction}
Among the intermediate valence semiconductors, TmSe attracts
particular interest because of its unique characteristics
\cite{Bucher75,Wachter94}. Unlike other members of the family,
both valence states of the rare earth counterpart, Tm$^{2+}$ and
Tm$^{3+}$ (the mean valence of Tm ion is $+2.75$ \cite{Bucher75}),
carry magnetic moments \cite{Wachter94}. Below the N{\'e}el
temperature $T_{\rm N}=3.5$~K, TmSe reveals a long-range ordered
antiferromagnetic phase with strongly pressure and magnetic field
dependent properties \cite{Guertin76}. These peculiarities are
considered to be the reason for a variety of unusual transport,
optical and magnetic properties of this intermediate valence
compound, like the magnetic field, temperature and pressure
dependence of the resistivity and Hall effect
\cite{Clayman77,Ribault80}, anisotropy of magnetization and
magnetostriction \cite{Nakanishi00}, temperature and pressure
dependence of quasi-elastic absorption in the neutron scattering
experiments \cite{Mignot00}, complicated structure of the optical
spectra in the low-energy region of a few meV
\cite{Batlogg81,Wachter94} and the origin of the insulating
antiferromagnetic state below $T_{\rm N}$
\cite{Bucher75,Batlogg79,Wachter94}. All of these properties are
likely to be related to the microscopic mechanism of the valence
fluctuations in TmSe which are under intensive study. It should be
noted, however, that to a significant extent the attention is
focused on the study of the magnetically ordered phase at $T <
T_{\rm N}$. At the same time also at higher temperatures, up to
the room temperature, TmSe reveals a behavior which is not
completely understood. For instance, the dc resistivity $\rho(T)$
is increasing towards liquid helium temperatures by about an order
of magnitude starting with 300~K and it does not show any sign of
an activated behavior as other mixed valence compounds do, like
YbB$_{12}$ \cite{Kasaya85,Moser85} or SmB$_6$
\cite{Menth69,Allen79}, where the opening of a hybridization gap
governs the resistivity behavior. The same is true for the Hall
measurements \cite{Clayman77,Andres78,Batlogg79}. It is assumed
that the Kondo scattering on the localized $f$-spins is
responsible for the temperature variation of resistivity and Hall
constant \cite{Clayman77,Haen87}, however, the correspondent
$\rho(T) \propto -\log T$ behavior is observed only in a narrow
temperature range between 4~K to 40~K \cite{Batlogg79}. The
magnetic susceptibility follows the Curie-Weiss law starting from
300~K down to 35~K \cite{Batlogg79}. Below this temperature the
magnetic moment of the Tm ions is enhanced, contrary to other
mixed valence semiconductors.

In earlier optical studies, a typical metallic reflectivity plasma
edge was observed at energies of a few eV
\cite{Wachter94,Batlogg81}. At lower energies signatures of a
non-Drude-like behavior were seen already at room temperature
\cite{Batlogg81}. The optical properties were also investigated in
the antiferromagnetic ground state \cite{Wachter94}. However, up
to now there exists no study of the temperature dependent
evolution of the infrared spectra in the paramagnetic phase. Here,
we present a detailed investigation of TmSe at temperatures 5~K to
300~K by means of far-infrared spectroscopic and dc resistivity
measurements. The aim of the present work was a comprehensive
study of the charge carrier dynamical properties in TmSe in the
paramagnetic phase. All experiments were performed in zero
magnetic field.

\section{Experiment and results}
We used high quality single crystals grown as described in
\cite{Matsumura98}. The typical size of the copper red single
crystals used for the optical experiments was 2 by 2 mm$^{2}$.
Reflectivity spectra were measured at nearly normal incidence
employing  a Bruker IFS 113v spectrometer in the frequency range
20 cm$^{-1} < \nu  < 10000$ cm$^{-1}$ at temperatures 5 K $< T <$
300 K. Reference spectra for calculating absolute values of
reflectivity were recorded using the in-situ Au-coating technique
\cite{Homes92}. In addition, quasi-optical reflectivity
measurements \cite{KozlovVolkov98} were performed down to lowest
frequencies $\nu =10$~cm$^{-1}$. From the frequency dependent
reflectivity we derived the real part of the optical conductivity
$\sigma_1$ and the permittivity $\epsilon_1$ by Kramers-Kronig
analysis. We used our dc resistivity data for the low-frequency
extrapolation and previously published data \cite{Batlogg81} for
the high frequency extrapolation. The dc resistivity was measured
using standard four-probe technique.

\begin{figure}[thb]
\begin{center}
{\scalebox{0.65}{\includegraphics*{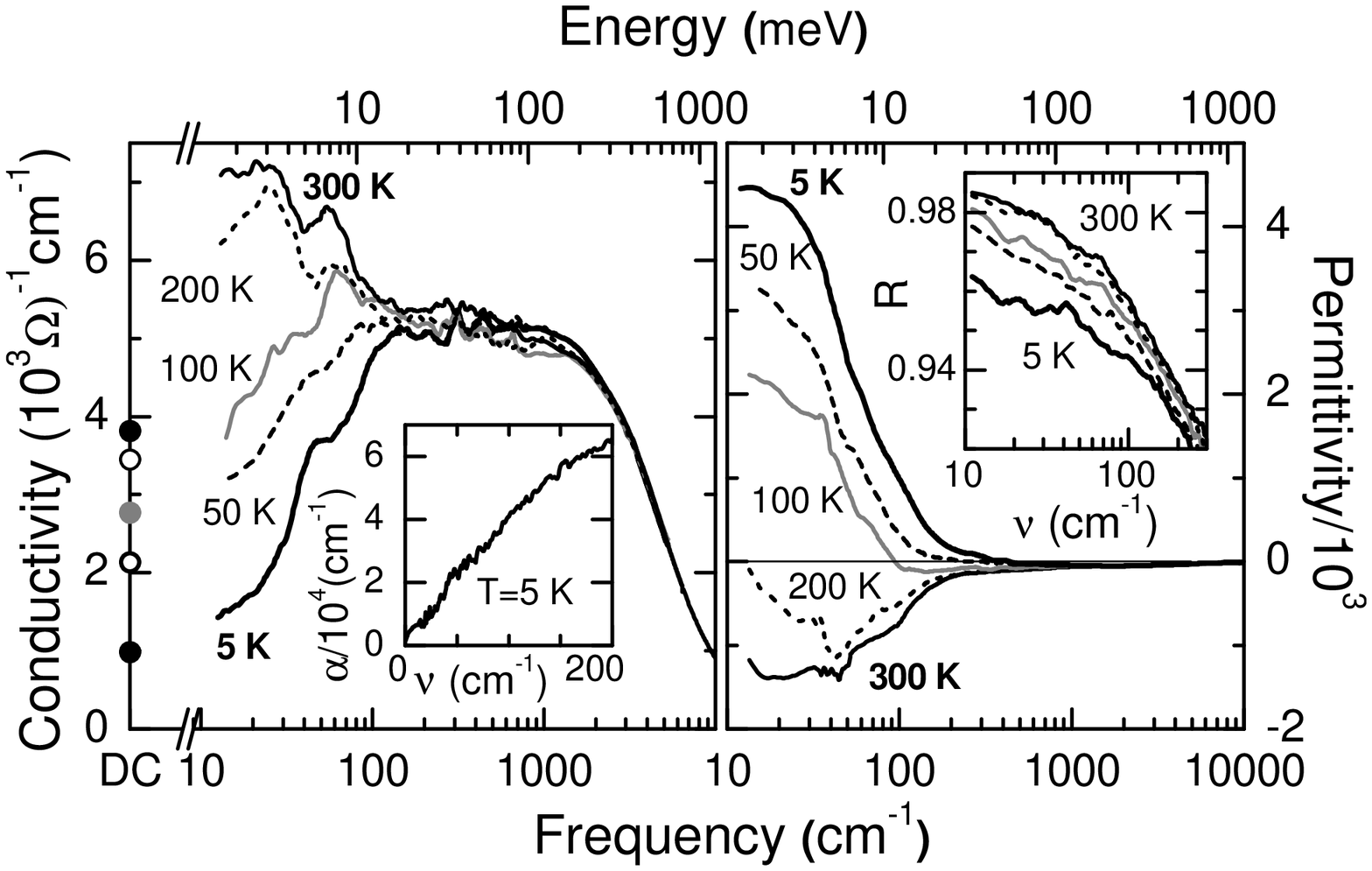}}}
\end{center}
\caption{\label{fig:1}Spectra of optical conductivity and
permittivity of TmSe at different temperatures as indicated. The
symbols on the conductivity ordinate represent the corresponding
dc values. In the insets the absorption coefficient $\alpha$ at 5
K (left panel) and the reflectivity $R$ (right panel) are shown.
The temperatures for the $\sigma_{\rm dc}$ and $R$ data are from
top to bottom 300 K, 200 K, 100K, 50 K, 5 K.}
\end{figure}

At all temperatures, the reflectivity of TmSe is high as usually
observed in conductors. The spectra of conductivity and
permittivity of TmSe derived from the recorded reflectivity data
are shown in fig.~\ref{fig:1} for several temperatures.
At frequencies above approximately 1000 cm$^{-1}$ the present
results are the same as reported previously at $T =$ 3.5 K and 300
K \cite{Wachter94,Batlogg81}. The spectra of conductivity and
dielectric constant are Drude-like  and almost temperature
independent in the mid- and near-infrared frequency region with a
characteristic roll-off around the scattering rate $\gamma$. Fits
of $\sigma(\nu)$ and $\epsilon(\nu)$ by the Drude model
\cite{DresselGruner02} yield $\gamma = 4700$~cm$^{-1}$ and the
plasma frequency $\nu_p =38500$~cm$^{-1}$ of the conduction
carriers; the corresponding plasma energy of 4.8~eV agrees with
the data given in \cite{Batlogg81}.

However, strong deviations from the Drude behavior are observed at
low frequencies, particularly below 100~cm$^{-1}$. At 300 K, an
additional bump is found in the conductivity spectrum in this
frequency region, coming along with a correspondent dispersion of
the permittivity. When cooling down to $T=200$~K, the spectra do
not change significantly. At even lower temperatures, however, the
low-frequency bump looses its strength and diminishes; the
reflectivity starts to decrease gradually. Cooling below 50~K, the
bump in $\sigma(\nu)$ turns into a gap-like feature. The spectra
do not change any further between $T=20$~K and 5~K.
Correspondingly, at low temperatures the permittivity is as large
as $\epsilon_1=4500$ below $\nu = $20~cm$^{-1}$.

\begin{figure}[t]
\begin{center}
{\scalebox{0.53}{\includegraphics*{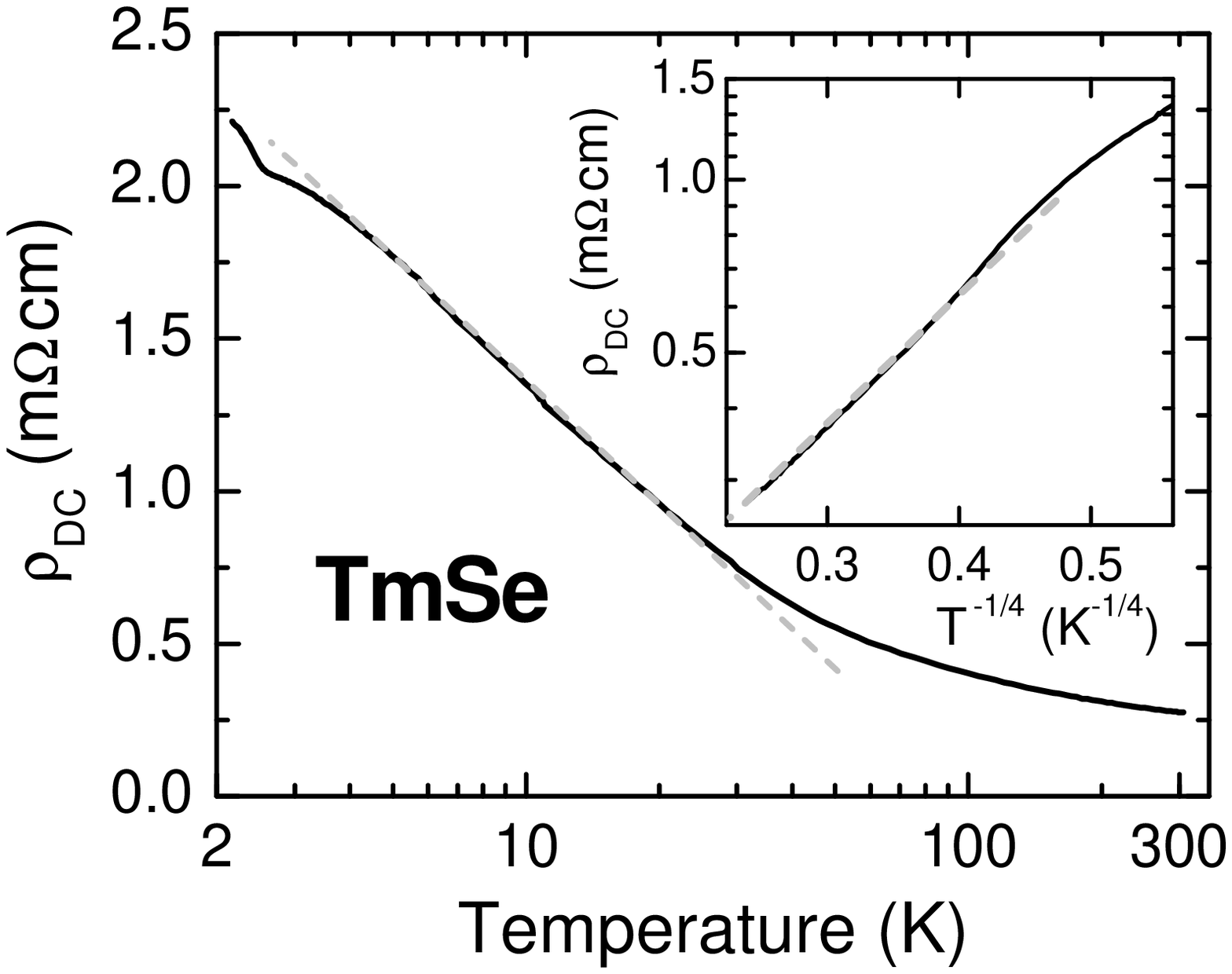}}}
\end{center}
\caption{\label{fig:2}Temperature dependence of the dc resistivity
of TmSe. The dashed line shows a Kondo behavior $\rho(T) \propto
-\log T$ above the N{\'e}el temperature $T_{\rm N}=3.5$~K and
below the Kondo temperature $T_{\rm K}=40$~K. The inset exhibits
the resistivity $\rho(T)$ {\it vs} $T^{-1/4}$; this temperature
dependence indicates variable-range hopping in the range $T = 40$
- 300~K.}
\end{figure}

The temperature dependent resistivity (fig.~\ref{fig:2}) is in
accord with the optical data and previously reported results
\cite{Clayman77,Batlogg79,Holtzberg85}. There is a very small
increase of $\rho(T)$ when cooling from 300 K to 200 K. Below
about 100~K, the resistance rises stronger and follows a
Kondo-type behavior $\rho(T)\propto -\log T$ between $T=35$~K and
the N{\'e}el temperature $T_{\rm N}$ \cite{Batlogg79}; hence the
Kondo temperature is estimated to be $T_{\rm K} \approx 40$~K
\cite{fn1}. Below 4~K $\rho(T)$ seems to saturate, but a sharp
kink followed by a steep increase is observed below $T_{\rm
N}=3.5$~K when the system enters the antiferromagnetic phase
\cite{Batlogg79}. Similar to \cite{Batlogg81}, the dc conductivity
$\sigma_{\rm dc}$ is noticeably
lower than the far-infrared values at elevated temperatures
($T=300$~K and 200~K), while the difference diminishes below 50~K.

\section{Discussion}
The charge dynamics of TmSe has been subject of discussion since
the transport characteristics like resistivity and Hall constant
are not thermally activated
\cite{Wachter94,Clayman77,Batlogg79,Haen80} and no low-temperature
decrease of magnetic susceptibility was reported
\cite{Batlogg79,Holzberg79}. These findings are in contrast to
intermediate-valence compounds with even number of electrons per
unit cell like YbB$_{12}$ and SmB$_6$. Here, signatures of
thermally activated behavior are observed in resistivity
\cite{Kasaya85,Menth69}, Hall \cite{Kasaya85,Allen79}  and optical
conductivity \cite{Travaglini84} which is in agreement with an
opening of a hybridization gap at the Fermi level in the density
of states upon cooling \cite{Wachter94}. The gap-like feature
observed at low temperatures in our optical spectra of TmSe,
however, cannot be connected to a hybridization gap in the density
of states for the following reasons:

(i) TmSe is an odd-number electron system for $T>T_{\rm N}$ which
should not exhibit a hybridization gap at $E_{\rm F}$ according to
the Luttinger theorem \cite{Luttinger60}. Only in the
antiferromagnetically ordered phase the hybridization gap is
`allowed', since the magnetic unit cell doubles and therefore
hosts an even number of $f$ and $d$ electrons. Indeed, a gap of
$\Delta_{\rm hyb} \approx 1 - 2$ meV was observed for $T<T_{\rm
N}$ \cite{Guentherodt82,Wachter94}. The periodic Anderson model
which has been treated with dynamical mean-field theory predicts
that the hybridization gap is expected to form below $T^{\ast} =
\Delta_{\rm hyb}/5$ \cite{Rozenberg96}. Interestingly, in TmSe
$T^{\ast}$ coincides with $T_{\rm N}$.

(ii) As seen from the inset of fig.~\ref{fig:1}, the absorption
coefficient $\alpha=2 k \omega/c$ ($k$ is the extinction
coefficient and $c$ the speed of the light) smoothly approaches
zero as $\nu \rightarrow 0$ and does not reveal an absorption-edge
structure, as observed in semiconductors with a gap in their
density of states \cite{Moss54}.

(iii) The resistivity is not thermally activated, instead it shows
a  $\rho \propto -\log T$ dependence.

\begin{figure}[b]
\begin{center}
{\scalebox{0.65}{\includegraphics*{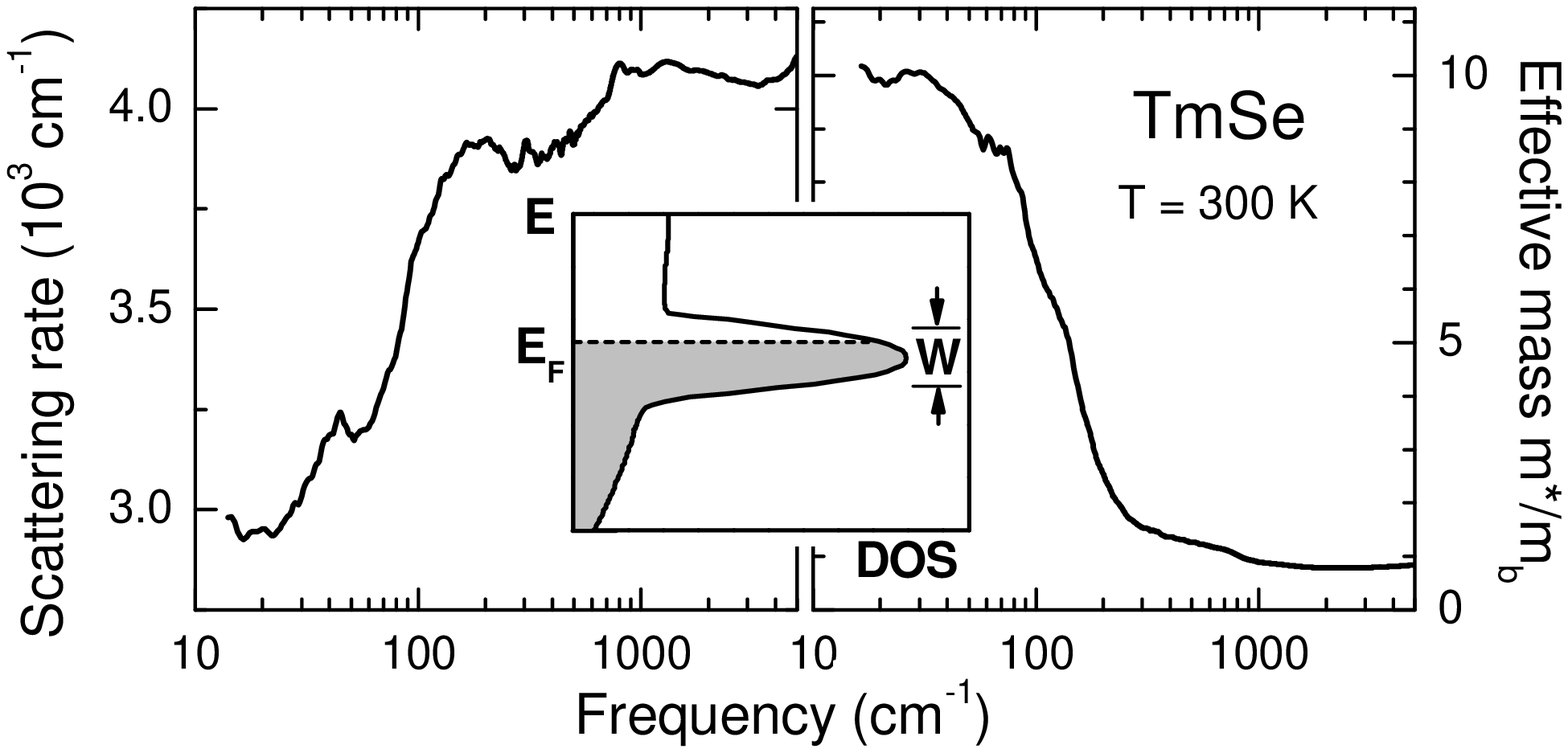}}} \end{center}
\caption{\label{fig:3}Frequency dependence of effective scattering
rate $\gamma(\nu)$ and effective mass $m^*(\nu)$ in TmSe at
ambient temperatures. The inset schematically shows the
quasi-particle density of states for a Kondo lattice system with
odd number of $f$ and $d$ electrons above the coherence
temperature $T^{\ast}$ with the Kondo resonance of width $W$ at
the Fermi level (after \cite{Varma76}).}
\end{figure}

Therefore, we can conclude that at $T >T_{\rm N}$ the charge
dynamics is dominated by inelastic Kondo scattering. In this case,
the Kondo lattice system TmSe behaves analogously to a dilute
Kondo system. The density of electronic states of such a system is
displayed in the inset of fig.~\ref{fig:3}. The bottom of the $5d$
conduction band overlaps the occupied $4f^{13}$ level of Tm$^{2+}$
where the electrons are localized. As a result a hybridized
$4f^{13}$-4$f^{12}5d$ state evolves at the Fermi level $E_{\rm
F}$, and the combined quasiparticle density of states reveals a
Kondo peak of width $W$ at $E_{\rm F}$. No hybridization gap is
present at $E_{\rm F}$ due to the reasons given above.

The optical and resistivity data can now be interpreted based on
this scheme. The density of states exhibits a peak at $E_{\rm F}$
well above the Kondo temperature, even at room temperature; this
fact was also observed for other heavy fermion and intermediate
valence compounds \cite{Wachter94,Bucher94}. Due to hybridization
the electrons in the vicinity of the Fermi level have both $d$ as
well as $f$ character.  They contribute to the optical response if
the electromagnetic radiation probes the region of enhanced
density of states inside the Kondo peak. Thus, the enhanced
conductivity and negative permittivity below 100 cm$^{-1}$
(fig.~\ref{fig:1}) at moderate temperatures $T=200$ - 300~K
indicate the formation of a Kondo peak already at such high
temperatures.

For a more detailed analysis of the frequency dependence of the
optical data we introduce a frequency dependent quasi-particle
scattering rate $\tau^{-1}(\omega)$ and effective mass
$m^*(\omega)$. Now, the low-frequency resonance is attributed to
the renormalization of $\tau^{-1}(\omega)$ and $m^*(\omega)$. Such
an analysis is widely applied for highly correlated electron
systems since it uses just one type of charge carriers and
therefore does not require to arbitrarily separate the response
into different components \cite{Degiorgi99}. The renormalization
is quantified within the generalized Drude model where a complex
scattering rate
$\hat{\Gamma}(\omega)=\tau^{-1}(\omega)-i\omega\lambda(\omega)$ is
introduced into the standard Drude formula:
\begin{equation}
\sigma_1(\omega)=\frac{(\omega_p)^2}{4\pi}
\frac{1}{\hat{\Gamma}(\omega)-i\omega}=\frac{(\omega_p)^2}{4\pi}
\frac{1}{\tau^{-1}(\omega)-i\omega(m^*(\omega)/m )}
\label{eq:gen-drude2}
\end{equation}
$\lambda(\omega)$ relates the frequency dependent effective mass
$m^*(\omega)/m =1+\lambda(\omega)$ to the frequency dependence of
the effective scattering rate $\tau^{-1}(\omega)$ which are both
linked by the Kramers-Kronig integrals \cite{DresselGruner02}. By
rearranging eq.~(\ref{eq:gen-drude2}), expressions for
$\tau^{-1}(\omega)$ and $m^*(\omega)$ in terms of real and
imaginary parts of the conductivity, $\sigma_1(\omega)$ and
$\sigma_2(\omega)=\omega\epsilon_1/(4\pi)$, are obtained:
\begin{equation}
\tau^{-1}(\omega)=\frac{(\omega_p)^2}{4\pi}
\frac{\sigma_1(\omega)}{|\hat{\sigma}(\omega)|^2}\,\,\,,\quad
\frac{m^*(\omega)}{m}=\frac{(\omega_p)^2}{4\pi}
\frac{\sigma_2(\omega)}{|\hat{\sigma}(\omega)|^2}\frac{1}{\omega}.
\end{equation}
Fig. \ref{fig:3} shows the calculated frequency dependences of
effective quasiparticles scattering rate $\gamma(\nu)
=\tau^{-1}(\nu)/(2\pi c)$ and effective mass $m^*(\nu)/m_b$ where
$m_b$ is the band mass and $\nu=\omega/(2\pi c)$.
At high frequencies the electrons in the conduction band dominate
with a scattering rate $\gamma \approx 4200$~cm$^{-1}$, a value
which is comparable to the one obtained from the simple Drude fit
(see above and fig.~\ref{fig:1}). For frequencies below
approximately 200~cm$^{-1}$, the effective scattering rate
decreases which indicates the reduced scattering due to the
admixture of heavy $f$ electrons. The effective mass obtained at
high frequencies is close to the band mass, $m^*\approx m_b$,
which is about 1.6 times the free electron mass $m_0$
\cite{Batlogg81}. Going to lower frequencies, it grows up to $m^*
\approx 10 m_b \approx 16 m_0$; this is an indication of
hybridization effects and of the presence of the Kondo peak at the
Fermi level already at room temperature. This peak in the density
of states should have no influence on the optical response for
frequencies considerably larger than its width, $h\nu \gg W$.
Since no significant temperature dependence is observed in the
optical spectra above 100~cm$^{-1}$, the width $W$ of the Kondo
peak in TmSe can be estimated from our optical data to be around
10 meV at room temperature. The Kondo temperature $T_{\rm K}=40$ K
obtained from the dc resistivity provides an estimation for $W$ at
low temperatures via the relation $W \approx k_{\rm B}T_{\rm K}=
3.4$ meV \cite{Martin82}. In agreement with theoretical
calculations \cite{Rozenberg96}, there is significant broadening
of the Kondo peak at temperatures $T \gg T_{\rm K}$.

When the temperature is lowered, the Fermi function becomes
sharper around $E_{\rm F}$ and consequently the admixture of
$f$-type carriers grows.  For $T<T_{\rm K}$ this leads to a
logarithmic increase of the resistivity as displayed in
fig.~\ref{fig:2}, and which is characteristic for incoherent Kondo
scattering. The consequence is a reduced low-energy spectral
weight in the conductivity spectrum which resembles a gap-like
feature. Now it becomes clear that this feature is a mobility gap
\cite{Mott79} and should be associated with the energy needed to
delocalize the charge carriers bound to local Kondo singlets. As
discussed above, the real hybridization gap will only develop in
the magnetically ordered phase below $T_{\rm N}=3.5$~K. Its
measured value of $\Delta_{\rm hybr.} \approx 1-2$~meV is
considerably smaller than the gap-like depression present in our
far-infrared conductivity data. It is not intended to
quantitatively analyze our data in detail. The behavior observed
in the optical spectra of TmSe at low frequencies is in
qualitative agreement with results obtained by simple early models
\cite{Wachter94,Batlogg81,Allen78} and more rigorous models
\cite{Anders97, Degiorgi99}, also including the magnetic field
dependences of the TmSe transport properties.

Finally, a brief comment on the relation between $\sigma_{\rm dc}$
and far-infrared response. At low temperatures, the optical
conductivity can be smoothly extrapolated into the dc limit. Here,
both dc and ac responses are determined by the heavy $f$-electron
component since for $T<50$~K the thermal energy $k_{\rm B}T$ is in
the order of the width $W$ of the Kondo peak. Contrary, as seen in
fig.~\ref{fig:1}, at high temperatures there is an obvious
mismatch between both data sets: for $T=300$~K and 200~K, where
the reflectivity is extremely high in the low-frequency limit,
$\sigma_1(\nu)$ shows clear signatures of metallic Drude-like
behavior above $\nu =10$~cm$^{-1}$, i.e., the conductivity
decreases with increasing frequency. However, in order to match
the dc data, $\sigma_1(\nu)$ has to decrease by a factor of two
below $\nu =10$~cm$^{-1}$; an observation which is in accordance
with previously reported results at 300 K \cite{Batlogg81}. As
shown in the inset of fig.~\ref{fig:2} , from 40~K up to 300 K the
dc resistivity follows the behavior $\rho(T)=
\rho(0)\exp[T_0/T]^{1/4}$ which is characteristic for the variable
range hopping \cite{Mott79} in systems with localization.
According to Mott \cite{Mott79}, the low-frequency conductivity in
this regime shows a behavior of the type $\sigma_1(\nu,T) =
\sigma_{\rm dc}(T) + \sigma_{0} \nu^s$ with an exponent $s\approx
1$ and the constant $\sigma_{0}$, which is frequently observed in
disordered materials at radio frequencies \cite{Dyre00}. Also for
TmSe the frequency dependent transport is expected to follow this
behavior which is, however, restricted to frequencies well below
$\nu =10$~cm$^{-1}$ and therefore explains the mismatch of dc and
ac data in fig.~\ref{fig:1}.

\section{Conclusions}
The optical spectra of conductivity and permittivity of the
intermediate valence compound TmSe have been measured down to low
frequencies ($\nu \approx 10$~cm$^{-1}$) at temperatures 5 -
300~K, i.e.\ in the paramagnetic state. At high temperatures, the
Drude-like mid-infrared conductivity is associated with the $d$
electrons in the conduction band which have a plasma energy of
4.8~eV and an effective mass close to free electron mass $m_0$
while the enhanced very far-infrared conductivity is connected
with an admixture of $f$ electrons to the charge-carrier
condensate as a result of the hybridization of itinerant $d$ and
localized $f$ electrons. These carriers are characterized by a
reduced scattering rate $\gamma$ and an enhanced effective mass of
$m^*\approx 16m_0$ at around room temperatures, indicating
correlation effects. At low temperatures, $5~{\rm K}<T<50$~K, a
gap-like feature is observed in the conductivity spectra for
frequencies $\nu<100$~cm$^{-1}$ which is accounted for as a
mobility gap due to localization of $d$ electrons on local Kondo
singlets.

\acknowledgments We are grateful to D. Faltermeier and P. Haas for
their experimental support. The project was supported by the
Deutsche Forschungsgemeinschaft (DFG) and the program for
fundamental research of the Division of Physical Sciences, RAS,
"Problems of Radiophysics".


\begin{thebibliography}{0}


\bibitem{Bucher75}
 \Name{Bucher E. {\it et al.}}
 \REVIEW{Phys. Rev. B}{11}{1975}{500}.
\bibitem{Wachter94}
\Name{Wachter P.} \Book{Handbook on the Physics and Chemistry of
Rare Earths} \Editor{Gschneider, K. A. Jr. {\it et al.}} \Vol{19}
\Publ{North-Holland, Amsterdam} \Year{1994} \Page{177}.
\bibitem{Guertin76}
 \Name{Guertin R. P., Foner S. \and Missell F. P.}
 \REVIEW{Phys. Rev. Lett.}{37}{1976}{529};
\Name{Bjerrum M{\o}ller H., Shapiro S. M. \and Birgeneau R. J.}
 \REVIEW{Phys. Rev. Lett.}{39}{1977}{1021}.
\bibitem{Clayman77}
\Name{Clayman B. P., Ward R. W. \and Tidman J. P.}
 \REVIEW{Phys. Rev B}{16}{1977}{3734}.
\bibitem{Ribault80}
\Name{Ribault M. {\it et al.}}
 \REVIEW{Phys. Rev. Lett.}{45}{1980}{1295};
\Name{Ohashi M. {\it et al.}}
 \REVIEW{Physica B}{259-261}{1999}{326}.
\bibitem{Nakanishi00}
\Name{Nakanishi Y. {\it et al.}}
 \REVIEW{Physica B}{281-282}{2000}{595}.
\bibitem{Mignot00}
\Name{Mignot J.-M. {\it et al.}}
 \REVIEW{Physica B}{276-278}{2000}{756};
\Name{Schlottmann P.} \REVIEW{Phys. Rev. B}{29}{1984}{630};
\Name{Shapiro S. M. \and Grier B. H.} \REVIEW{Phys. Rev.
B}{25}{1982}{1457}.
\bibitem{Batlogg81}
\Name{Batlogg B.} \REVIEW{Phys. Rev. B}{23}{1981}{1827}.
\bibitem{Batlogg79}
\Name{Batlogg B. {\it et al.}} \REVIEW{Phys. Rev.
B}{19}{1979}{247}.
\bibitem{Kasaya85}
\Name{Kasaya M. {\it et al.}} \REVIEW{J. Magn. Magn.
Mater.}{47-48}{1985}{429}.
\bibitem{Moser85}
\Name{Moser M. {\it et al.}} \REVIEW{Sol. State
Commun.}{54}{1985}{241}.
\bibitem{Menth69}
\Name{Menth A., Buehler E. \and Geballe T. H.} \REVIEW{Phys. Rev.
Lett.}{22}{1969}{295}.
\bibitem{Allen79}
\Name{Allen J. W., Batlogg B. \and Wachter P.} \REVIEW{Phys. Rev.
B}{20}{1979}{4807}.
\bibitem{Andres78}
\Name{Andres K. {\it et al.}} \REVIEW{Sol. State
Commun.}{27}{1978}{825}.
\bibitem{Haen87}
\Name{Haen P. {\it et al.}} \REVIEW{J. Magn. Magn. Mater.
}{63-64}{1987}{603}.
\bibitem{Matsumura98}
\Name{Matsumura T. {\it et al.}} \REVIEW{J. Phys. Soc.
Jpn.}{67}{1998}{612}.
\bibitem{Homes92}
\Name{Homes C. C. {\it et al.}} \REVIEW{Appl.
Optics}{32}{1992}{2976}.
\bibitem{KozlovVolkov98}
\Name{Kozlov G.V. \and Volkov A.A.} \Book{Millimeter and
Submillimeter Wave Spectroscopy of Solids} \Editor{Gr{\"u}ner G.}
\Publ{Springer, Berlin} \Year{1998} \Page{51}.
\bibitem{DresselGruner02}
\Name{Dressel M. \and Gr{\"u}ner G.} \Book{Electrodynamics of
Solids} \Publ{Cambridge University Press, Cambridge} \Year{2002}.
\bibitem{Holtzberg85}
\Name{Holtzberg F. {\it et al.}} \REVIEW{J. Appl.
Phys.}{57}{1985}{3152}.
\bibitem{fn1}The Kondo temperature $T_{\rm K}$ usually marks the
minimum resistivity which separates the region with
$\rho(T)\propto -\log T$ behavior from a metallic temperature
dependence of $\rho(T)$ at higher $T$. Here, it marks the
transition from the Kondo behavior to hopping transport around 40
K.
\bibitem{Haen80}
\Name{Haen P. {\it et al.}} \REVIEW{J. Magn. Magn.
Mater.}{15-18}{1980}{989}.
\bibitem{Holzberg79}
\Name{Holtzberg F., Penney T. \and Tounier R.} \REVIEW{J. de
Physique}{40}{1979}{314}.
\bibitem{Travaglini84}
\Name{Travaglini G. \and Wachter P.} \REVIEW{Phys. Rev.
B}{29}{1984}{893}; \Name{Gorshunov B. {\it et al.}} \REVIEW{Phys.
Rev. B}{59}{1999}{1808}; \Name{Okamura H. {\it et al.}}
\REVIEW{Phys. Rev. B}{58}{1998}{R7496}.
\bibitem{Luttinger60}
\Name{Luttinger J. M.} \REVIEW{Phys. Rev.}{119}{1960}{1153};
\Name{Martin R. M. \and Allen J. W.} \REVIEW{ J. Appl.
Phys.}{50}{1979}{7561}.
\bibitem{Guentherodt82}
\Name{G\"{u}ntherodt G. {\it et al.}} \REVIEW{Phys. Rev.
Lett.}{49}{1982}{1030}.
\bibitem{Rozenberg96}
\Name{Rozenberg M. J., Kotliar G. \and Kajueter H.} \REVIEW{Phys.
Rev. B}{54}{1996}{8452}.
\bibitem{Varma76}
\Name{Varma C. M.} \REVIEW{Rev. Mod. Phys.}{48}{1976}{219}.
\bibitem{Moss54}
\Name{Moss T. S.} \REVIEW{Proc. Phys. Soc. B}{67}{1954}{775}.
\bibitem{Bucher94}
\Name{Bucher B. {\it et al.}} \REVIEW{Phys. Rev.
Lett.}{72}{1994}{522}.
\bibitem{Degiorgi99}
\Name{Degiorgi L.} \REVIEW{Rev. Mod. Phys.}{71}{1999}{687}.
\bibitem{Martin82}
\Name{Martin R. M.} \REVIEW{Phys. Rev. Lett.}{48}{1982}{362}.
\bibitem{Mott79}
\Name{Mott N. F. \and Davis E. A.} \Book{Electronic Processes in
Non-Crystalline Materials}, 2nd edition \Publ{Clarendon Press,
Oxford} \Year{1979}.
\bibitem{Allen78}
\Name{Allen J. W. {\it et al.}} \REVIEW{J. Appl.
Phys.}{49}{1978}{2078}.
\bibitem{Anders97}
\Name{Anders F. B., Jarrell M. \and Cox D. L.} \REVIEW{Phys. Rev.
Lett.}{78}{1997}{2000}.
\bibitem{Dyre00}
\Name{Dyre J. C. \and Schr{\o}der T. B.} \REVIEW{Rev. Mod.
Phys.}{72}{2000}{873}.
%
\end{thebibliography}
\end{document}